# A Comparative Study of Homomorphic and Searchable Encryption Schemes for Cloud Computing

Prasanna B T
Department of ISE
EPCET
Bengaluru-560049, INDIA
prasi.bt@gmail.com

C B Akki
Department of ISE
SJBIT
Bengaluru-560060, INDIA
akki.channappa@gmail.com

***Abstract*— Cloud computing is a popular distributed network and utility model based technology. Since in cloud the data is outsourced to third parties, the protection of confidentiality and privacy of user data becomes important. Different methods for securing the data in cloud have been proposed by researchers including but not limited to Oblivious RAM, Searchable Encryption, Functional Encryption, Homomorphic Encryption etc. This paper focuses on Searchable and Homomorphic Encryption methods. Finally, a comparative study of these two efficient cloud cryptographic methods has been carried out and given here.**

## I. SECURITY IN CLOUD COMPUTING

Cloud Computing is a distributed network meant for providing service by cloud provider to the consumers on rental basis [1]. Owners of data store their data in cloud which therefore need to be secured. By storing data in encrypted form, one can maintain the confidentiality and privacy of data in cloud. Many cryptographic methods have been devised to address the issue of confidentiality and privacy of owner's data in cloud. An in-depth survey has been done on cloud related security issues, challenges and cryptographic algorithms by Prasanna and Akki [2]. Among them homomorphic and searchable encryption methods are most popular, where one can perform computation and search on ciphertext without disclosing plaintext. The authors focus more on these two techniques in this paper.

The paper is organized as follows: Section 2 discusses criteria's used for developing Homomorphic Encryption methods. The section also briefs some of the known Fully Homomorphic Encryption methods and their corresponding technologies used. Section 3 focuses on related work on Searchable Encryption methods and different criteria's used to categorize them. Section 4 compares different known cloud cryptographic methods using efficiency and security as parameters. Section 5 concludes with an utter need of efficient cryptographic methods like Searchable Encryption for further study and research.

## II. HOMOMORPHIC ENCRYPTION

In cloud, the primary concern is of maintaining both confidentiality and privacy of owner's data from untrusted users. The concept of homomorphism introduced in 1978, by Rivest et al [3], can be used for securing the data stored in cloud from unauthorized users.

Two messages m1 and m2 are encrypted by using any known encryption method E with public or private key pk, where C1 and C2 are their corresponding ciphertexts (i.e. C1= $E_{pk}$ (m1) and C2= $E_{pk}$ (m2)). The HE scheme performs computation like addition, multiplication etc. between C1 and C2 without decryption. The obtained result is also in encrypted form. The general architecture of the HE scheme is shown in figure 1.

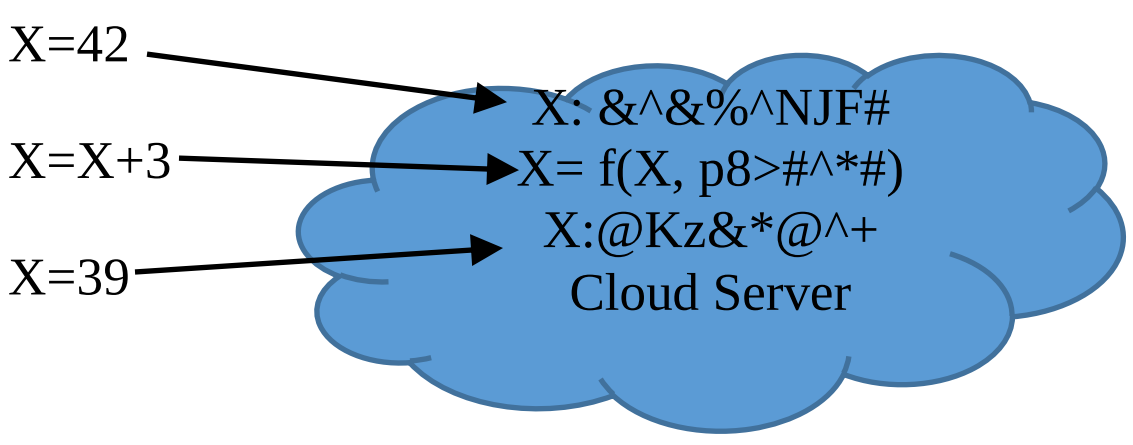

Figure 1. General architecture of Homomorphic Encryption

As shown in figure 1, user performs computation on encrypted data in cloud. Which in turn preserves the privacy and security of user data in cloud.

The HE algorithms are designed and developed using different mathematical constructs. Each HE algorithm is developed mainly based on two factors: Hardness of breaking the security of encryption and Efficiency (in terms of time) of execution of encryption, decryption processes. Numerous HE models have been proposed based on certain criteria.

### A. *Classification Of HE's:*

Based on keys used HE schemes can be symmetric, asymmetric and hybrid. Further HE methods can be broadly classified into Partial Homomorphic Encryption (PHE) and Fully Homomorphic Encryption (FHE). PHE performs only a limited number of operations for example either addition or multiplication on encrypted data. FHE performs both addition and subtraction operations on encrypted data many times. Based on ciphertext generated, an encryption scheme can be probabilistic or deterministic. In probabilistic encryption each time a plaintext encrypted, results in different ciphertext, whereas deterministic encryption, always results in the same ciphertext. To secure the encryption or decryption process, different mathematical constructs are used like composite residuocity, high degree residuocity and quadratic residuocity. Trapdoor functions are used to secure data in HE. They are one way functions used during encryption and decryption process. Different mathematical constructs are used to create trapdoors. In Malleable cryptosystems a given ciphertext can be transformed to another by adversary to get plaintext. Versatility of a HE scheme is its ability to support many circuits to do computation on encrypted data. Versatility should be high to support more operations on encrypted data. Efficiency of HE schemes is considered to be good if generated ciphertext size is small and time taken to run encryption, decryption or recryption process is less. Table 1, lists some of the criteria's considered during design and development of HE schemes.

Some of the known PHE methods include Unpadded RSA, ElGamal, Goldwasser-Micali, Benaloh, Paillier, Okamoto-Uchiyama, Naccache-Stern, Damgard-Jurik and Boneh-Goh-Nissim etc. Some of the known FHE methods are Gentry, Van Dijk, Smart-Vercauteren, Stehle-Steinfield, Ogura, Lyubashevsky, Gentry-Halevi, Brakerski-Vaikuntanathan, Brakerski-Gentry-Vaikuntanathan and Chunsheng etc. Prasanna and Akki [2] in their survey paper discussed different known PHE and FHE schemes along with their limitations and benefits.

Table 1: HE schemes criteria's and types

| Criteria | Types |
|---|---|
| Trapdoor functions | 1. Polynomial over finite field<br>2. Discrete logarithms over groups |
| HE schemes | 1. Partial homomorphic encryption.<br>2.Fully homomorphic encryption |
| Encryption method | 1. Probabilistic encryption.<br>2.Deterministic encryption |
| Hardness of encryption | 1. Composite Residuosity.<br>2. High degree Residuosity.<br>3. Quadratic Residuosity |
| Malleability | 1. Malleable.<br>2. Non Malleable |
| Versatility | 1. Versatile.<br>2. Non Versatile |
| Efficiency based on computation speed | 1. Encryption.<br>2. Decryption<br>3. Recryption |
| Efficiency based on ciphertext size | 1. Small ciphertext size<br>2. Large ciphertext size |
| Practical or not | 1. Practically implementable.<br>2.Practically complex to implement |
| Cryptosystem | 1. Symmetric.<br>2. Asymmetric.<br>3. Hybrid |

### B. *Preliminaries (HE):*

The mathematical terms used in Table 2 and Table 3 are defined here:

**Ideal lattices:** Lattices were first studied by mathematicians Joseph Louis Lagrange and Carl Friedrich Gauss. In 1996, Miklos Ajtai and Micciancio [4] discussed the use of lattices as cryptography primitive. Micciancio defined lattices as general class of cyclic lattices.

A lattice L is a set of points in the n-dimensional Euclidean space $R^n$ with a strong property of periodicity. A basis of L is a set of vectors such that any element of L is uniquely represented as their linear combination with integer coefficients. When n is at least 2, each lattice has infinitely many different bases. All lattices over $R^n$ have infinitely many elements, whereas in cryptography entities such as the ciphertext, public key, and private key must be taken from a finite space (bit strings of some fixed length). Therefore the lattices used for cryptography are actually lattices over a finite field.

There are two types of lattice based mathematical problems. They are the Shortest Vector Problem (SVP) and the Closest Vector Problem (CVP). SVP: Given a basis of a lattice, find the shortest vector in the lattice. CVP: Given a basis of a lattice and a vector not in the lattice, find the lattice vector with the least distance to the first vector. Lattice based cryptography refers to any system whose security depends on computational assumptions based on lattices.

**Approximate GCD:** Approximate GCD is a difficult problem of symbolic-numeric computation. Some of the algorithms try to guess the near approximate solution of the given problem, later the accuracy and precision of the obtained approximate solution is improved. The approximate GCD is used to find one or more divisors which is the greatest common divisor of the approximate numbers a and b of two given numbers a0 and b0.

**Algebraic number theory:** Algebraic number theory is a branch of number theory that studies algebraic structures related to algebraic integers. This is generally accomplished by considering a ring of algebraic integers O in an algebraic number field, and studying their algebraic properties such as factorization, the behavior of ideals, and field extensions.

**Sparse Subset Sum Problem:** The subset sum problem is an important problem in complexity theory and cryptography. The problem deals with a given set of integers, and finding a non-empty subset whose sum is zero. For example, given the set {−7, −3, −2, 5, 8}, the answer is yes because the subset {−3, −2, 5} sums to zero. The problem is NP-complete.

**Learning with Error (LWE):** The LWE is a generalization of the parity learning problem. The LWE problem is to distinguish random linear equations, which has been perturbed by a small amount of noise, from truly uniform ones. The problem has been shown to be as hard as worst-case lattice problems.

An algorithm is said to solve the LWE problem if, when given access to samples $(x.y)$ where $x \in Z_q^n$ and $y \in Z_q$ with the assurance, for some fixed linear function $f : Z_q^n \rightarrow Z_q$, that $y = f(x)$ with high probability and deviates from it according to some known noise mode, the algorithm can recreate $f$ or some close approximation of it with high probability. Where n is positive integer, q is odd prime and f is a linear function.

**Ring Learning with Errors (RLWE):** For a polynomial ring $R_q = Z_q[x]/(f(x))$ and a random polynomial $w \in R_q$, it is computationally hard to distinguish the uniform distribution over $R_q\, X\, R_q$ from ordered pairs of the form $(a_i, a_i w + e_i)$, where $a_i$ are uniformly distributed in $R_q$ and $e_i$ are polynomials in $R$ whose coefficients are independently distributed Gaussians**.**

**Nth degree Truncated polynomial ring (NTRU):** The NTRU Encrypt public key cryptosystem, also known as the NTRU encryption algorithm, is a lattice-based alternative to RSA and ECC (Elliptic Curve Cryptosystem) and is based on the shortest vector problem in a lattice.

### C. *Fully Homomorphic Encryptions(FHE):*

In 2009, Gentry for the first time introduced the concept of FHE. After that many researchers contributed towards making FHE practical.

Some of the known FHE methods and their corresponding technologies involved are listed in Table 2. Table 3 compares different FHE schemes based on LWE technique.

Size of different parameters mentioned in Table 3 is almost same. The BGV and Vercauteren schemes are based on RLWE, and uses less number of bits compared to BV and Brakerski schemes. Vercauteren scheme uses less number of bits for secret and public key in turn increases the processing speed of homomorphic encryption.

The main drawback of using HE schemes in cloud is its complex mathematical constructs involved. Encryption, decryption and recryption processes involving these mathematical constructs take more time for processing. Researchers are working towards improving the security of HE schemes while maintaining efficiency (in terms of time) of these algorithms.

## III. SEARCHABLE ENCRYPTION

Homomorphic Encryption schemes have been surveyed so far to address the issue of privacy and confidentiality. The same issues can also be addressed by using one more efficient cryptographic tool called Searchable Encryption. The general architecture of SE schemes is shown in figure 2.

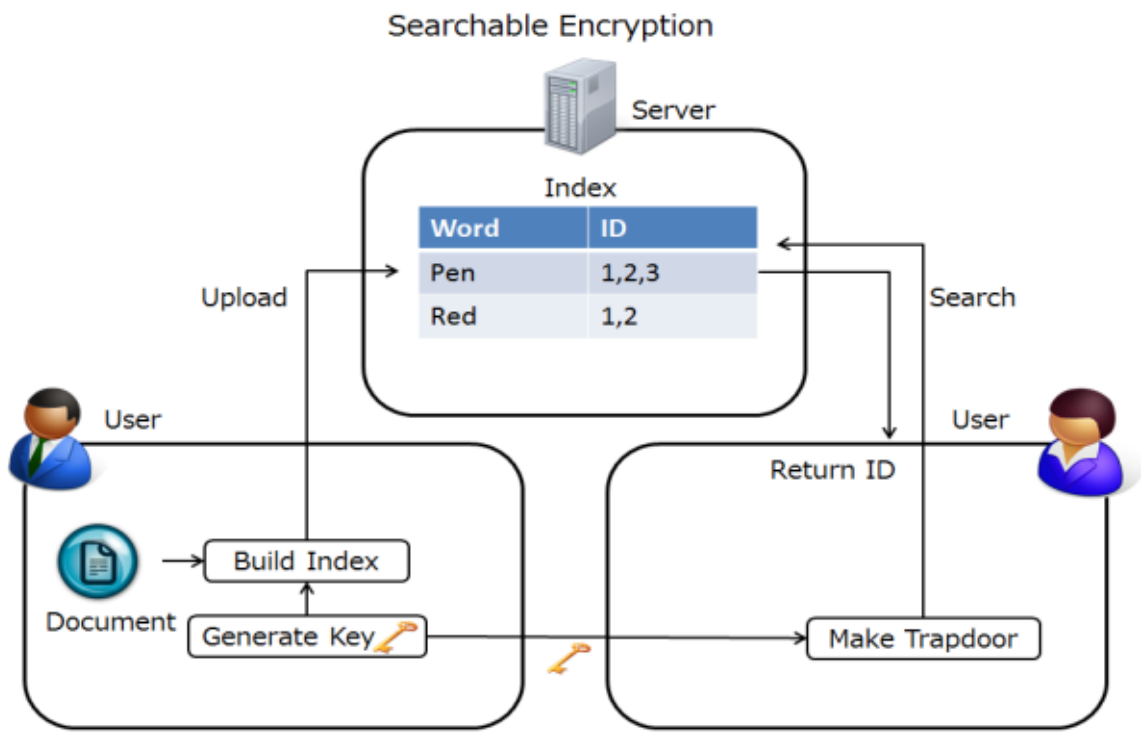

Figure 2. General Architecture of SE Schemes

As shown in figure 2, data owner encrypts the document and correponding index fils and uploads the same into cloud. In cloud, the keywords(trapdoor)sent by clients is used for search and retrived encrypted document is sent to the clients.

SE maintains the confidentiality and privacy of owner's data by facilitating searching keywords directly on encrypted data. Users can upload their encrypted data to cloud. Later, the authorized users can perform private keyword search on encrypted data in cloud. Multiple domains like cryptography, indexing, storage etc. are involved in devising efficient, secure, SE algorithms over encrypted files. The participants of a secure search model in a cloud, typically involves data owner, data user and cloud server. Data owner encrypts the files and corresponding keywords based index files by using any known cryptographic algorithms. Both the encrypted files and index files are uploaded to the cloud server. To search for keywords data user requests trapdoors from data owner or generates trapdoors himself/herself. These trapdoors are used to search encrypted files in cloud server. Cloud server searches for the keywords in database and encrypted search results are returned to the data user. Many researchers are involved in designing and developing SE algorithms with having high efficiency (in terms of time).

Table 2: FHE Schemes and Techniques Used

| FHE Schemes | Year | Based on |
|---|---|---|
| Gentry | 2009 | Ideal Lattice |
| Van Dijk et al (DGHV) | 2010 | Approximate GCD |
| Smart-Vercauteren | 2010 | Elementary theory of algebraic number fields |
| Stehle-Steinfield | 2010 | Sparse Subset Sum Problem (SSSP) |
| Ogura et al | 2010 | Relation between Circuit depth and Eigen Values |
| Lyubashevsky et al | 2010 | Ideal lattice |
| Gentry-Halevi | 2011 | Ideal Lattice Implementation |
| Brakerski-Vaikuntanathan | 2011 | Learning with Error (LWE) |
| Brakerski-Gentry-Vaikuntanathan (BGV) | 2011 | Ring Learning with Error (RLWE) |
| Lyubashevsky et al | 2012 | Ring Learning with Error (RLWE) |
| Chunsheng G | 2012 | Hardness of factoring integer<br>Solving Diophantine equation problem<br>Finding Approximate GCD problem |
| Jean-Sebastien et al | 2013 | Batch FHE over the integers |
| Kurt Rohloff et al | 2014 | Nth degree Truncated polynomial Ring (NTRU) |

Table 3: Comparison of FHE Schemes Based on LWE

| Scheme | Based on Technique | Secret key size | Public key size | Ciphertext size |
|---|---|---|---|---|
| BV | LWE | n log q | $O(n^2 \log^2 q)$ | (n+1) log q |
| BGV | LWE and RLWE | 2d log q | 2dn log q | 2d log q |
| Brakerski | LWE | n log q | $O(n^2 \log^2 q)$ | (n+1) log q |
| Vercauteren | RLWE | d | 2d log q | 2d log q |

*Where p and q is prime number. n=pxq and d is length in bits.

## A. *Related Work:*

Song et al. [5] for the first time proposed practical symmetric searchable encryption method. In this scheme the file is encrypted word by word. To search for a keyword user sends the keyword with same key to the cloud. The drawback of this scheme is that the word frequency will be revealed. Goh [6] tried to overcome the drawback of Song's scheme by constructing secure index table using pseudorandom functions and unique document identifier randomized bloom filters. Bosch et al. [18] worked on the concept given by Goh et al. and introduced the concept of wild card searches. The drawback of this scheme is that bloom filters may introduce false positives. In Chang's et al. [7] proposed scheme, an index is built for each document. The scheme is more secured compared to Goh's scheme since number of words in a file is not disclosed. The limitation of this scheme is that it is less efficient and does not support arbitrary updates with new words. Golle et al. [9] scheme allows multiple keyword searches with one encrypted query. But this scheme is not practical. Curtmola et al.[8] for the first time proposed the concept of symmetric searchable encryption (SSE), later on Kamara et al.[23] proposed an extended version of SSE called dynamic SSE(DSSE), where addition and deletion of documents can be performed in index table. Curtmola's scheme and Kamara's scheme are limited to keyword searches, and range queries are not supported. Hacigumus et al. [25] proposed an SQL query support by using bucketization technique. The drawback is with respect to bucketization technique where the scheme does not support scalability. Hore et al. [19] proposed a scheme which applies bucketization to multi-dimensional data that also supports range queries. The limitation is same as in Hacigumus scheme, which is of scalability. Order Preserving Encryption (OPE) based schemes supports range queries. The drawback with the schemes supporting OPE is of revealing the order relation between ciphertext data. Popa et al. [26] proposed a technique called CryptDB to protect databases. CryptDB uses proxy server as an interface to the database. Uses range queries and homomorphic encryption methods like Paillier scheme for searching and performing computations on encrypted data respectively. The drawback is of trusting the proxy server. All these schemes are based on single user [24].

The first public key encryption with keyword search (PEKS) was proposed by Boneh et al. [15]. The scheme suffers from inference attack on trapdoor encryption method. Baek et al. [13], Rhee et al. [20] improved hardness of security of Boneh's scheme. Baek's scheme introduces the concept of conjunction of keyword search. The public key encryption methods are computationally time consuming and complex that makes these algorithms inefficient. In Yang et al. [22] scheme the encrypted data is searched by individual users using a unique key allotted to them. The scheme suffers from key management. Boneh et al. [16], [17] discussed functional encryption and related to conjunctional search, range queries and subset queries. Katz et al. [12] scheme is an updated version of Boneh's scheme [16] and discussed predicate encryption for inner products and supports both conjunctions and disjunctions search on encrypted data. Shen et al. [21] scheme is based on symmetric key and uses predicate encryption. Li et al. [28] scheme was based on Hidden Vector Encryption (HVE) that uses multiple trusted agencies to distribute search capacities to users. The drawback of these schemes is key management and complex methodologies involved. Cao et al. [11] scheme uses multi keyword search technique. Here the database owner builds an index table which is very expensive and cannot be changed dynamically. All these schemes are based on semi fledged multiuser [24].

Hwang et.al [10] proposed public key encryption with conjunctive keyword search (PECK) where multi users can be involved. Here the ciphertext size increases linearly with the number of users. Bao et al. [14] proposed a scheme based on bilinear maps, for multiuser setting. The drawback is computational complexity involving bilinear maps. Dong et al. [27] proposed a scheme based on Symmetric Data Encryption (SDE) and depends on proxy encryption. This scheme is efficient in performing search on encrypted data. Shao et al. proposed proxy re-encryption with keyword search. This scheme involves managing keys that makes this scheme inefficient. Muhammad Rizwan Asghar et al. [24] propsed a scheme based on full-fledged multi user settings and complex query search. This scheme supports SQL like queries on encrypted databases. Here the queries are represented by using tree data structure. All these schemes are based on full-fledged multiuser [24].

Many scientists across the globe designed numerous SE methods. They considered different criteria's to develop SE schemes. The different criteria's considered to classify the SE schemes are discussed in the following sections.

*B. Preliminaries(SE)::*

The following section defines technical terms used in Tables 4, 5, 6, 7, 8, 9 and 10.

**ID based Encryption (IBE):** ID-based encryption (or Identity-Based Encryption (IBE)) is an important primitive of cryptography. The public key contains some unique identities of each user hence the name IDE. For example the key generation function can use the text-value of the name or domain name or the physical IP address to generate a public key.

**Edit distance:** The edit distance between two words w1 and w2 is the number of operations required to transform one of them into the other. The three primitive operations are 1) Substitution: changing one character to another in a word; 2) Deletion: deleting one character from a word; 3) Insertion: inserting a single character into a word.

**Bedtree**: All-purpose edit distance based indexing scheme using the B+ tree structure.

**Predicate Encryption (PE):** In PE schemes each ciphertext C is associated with a binary attribute vector a = (a1... an) and keys K are associated with predicates. A key K can decrypt a ciphertext C if and only if the attribute vector of the ciphertext satisfies the predicate of the key. Predicate encryption schemes can be used to implement fine-grained access control on encrypted data and to perform search on encrypted data.

**Hidden Vector Encryption (HVE):** HVE is predicate encryption where in each ciphertext C is associated with a binary vector a = (a1... an) and each key K is associated with binary vector b = (b1... bn) with "don't care" entries. Key K can decrypt ciphertext C if and only if a, b agree for all i for which bi is not "don't care" entries.

**Diffie-Hellman Problem (DHP):** Whitfield Diffie and Martin Hellman proposed a mathematical problem used in cryptography. DHP is used to encrypt a message easily, but reversing the same encryption is difficult.

The problem statement: "Given an element $g$ and the values of $g^x$ and $g^y$, what is the value of $g^{xy}$?" where g is a generator of some group and x and y are randomly chosen integers.

**Decisional Diffie-Hellman (DDH):** The decisional Diffie–Hellman (DDH) assumption is a computational hardness assumption about a certain problem involving discrete logarithms in cyclic groups. It is used as the basis to prove the security of many cryptographic protocols, most notably the ElGamal and Cramer–Shoup cryptosystems. The problem statement: Given $g^x$ and $g^y$ for uniformly and independently chosen x, y random integers**.** The value of $g^{xy}$ "looks like" a random element in G. where g is a generator and G is a multiplicative cyclic group.

**Boolean Keyword Search:** Boolean keyword search combines keywords with operators like AND, OR, NOT to get a result. For example the keyword Bangalore AND Mangalore limits the search to the documents that contain only these two words.

**Fuzzy keyword search algorithms:** The Three Searching Schemas for Fuzzy Keyword Search over Encrypted cloud data are 1.Wildcard Based technique 2. Gram based technique 3. Tree traverse search scheme.

**Wildcard –Based Technique:** This Technique uses fuzzy set edits distance to solve the problems. The Edit distance can be substitution, Deletion and Insertion. By using the Wildcard-Based Technique, the Fuzzy keyword search can be more efficient because different edit distance techniques are used which can be used to find for finding the keywords.

**Gram-Based Technique:** Another efficient technique for constructing fuzzy set is based on grams. The gram of a string is a substring that can be used as a signature for efficient approximate search. Gram has been widely used for constructing Inverted List for approximate string search.

**Tree Traverse Search Technique:** The search efficiency, tree -traverse search scheme, where a multi-way tree is constructed for storing the fuzzy keyword set over a finite symbol set. The key idea behind this construction is that all trapdoors sharing a common prefix may have common nodes. All fuzzy words in the tree can be found by a depth-first search.

**Index organized by keyword and inverted index**: This allows fast retrieval of files by using hashing technique. The search complexity is constant.

**Index constructed per document:** This uses linear search for number of documents in the database. Search time is more.

**Tree based index structure:** This uses tree data structure for search. Search time is considerably less compared to index constructed per document method.

**Full Domain Based (Linear):** One is to perform full-domain search, in which a search will sequentially go through every data item in order to test some criteria.

**Index Based:** The other is index-based search or keyword-based search, where every data item is firstly characterized by a list of keywords which are then used to build a search index for the data item. Later, when a search takes place, the criteria will be tested based on the indexes instead of the contents of data items.

Algorithms are broadly categorized into linear or index based searchable algorithms depending on search technique used.

### *C. Classification of SE's:*

Based on data retrieval, query numbers and their type, result, number of participants, security proof models, performance, encryption models, search types, evaluation parameters etc. SE algorithms have been designed and categorized. Table 4 lists some of the known criteria's used to design SE methods.

Two methods of SE found in the literature i.e. private and public key methods. In private key method the same key is used for both encryption and keyword search. In public key scheme different keys are used for encryption and keyword search. Public key algorithms are further classified based on keyword search types like conjunctive keyword search, fuzzy keyword search etc. The general architecture of SE is shown in figure 1.

Table 4: Criteria's for choosing SE schemes

| Criteria | Types |
|---|---|
| Factors | When searching, what must be protected 1.Retrieved Data: The data in cloud (storage). 2. Search Query: What to search? 3. Result: Outcome of the query. |
| Scenario | 1. Single vs Multiple Queries. 2. Adaptive Queries: Query depends on previous result. 3. Non Adaptive Queries: Multiple Queries, but not related. |
| Number of participants | 1. Single user can query. 2. Multiple users can query. |
| Security proof models | 1. Random Oracle model. 2. Standard model. |
| Number of keywords used for searching | 1. Single Keyword. 2. Multiple keywords. |
| Encryption techniques | 1. Symmetric encryption. 2. Asymmetric or public key encryption. |
| Searching techniques | 1. Searchable symmetric key. 2. Searchable public key. 3. Secure Index. 4. Privacy preserving keyword search 5. Privacy preserving multi keyword ranked search: k-nearest neighbor technique. 6. Inner product similarity: Number of Query keywords appearing in a document. 7. Conjunctive Search. 8. Disjunctive Search. 9. Single keyword searchable encryption. 10. Fuzzy searchable encryption. 11. Boolean keyword searchable encryption. |
| Performance | 1. Number of retrieved documents vs precision. 2. Number of retrieved documents vs privacy. 3. Number of docs vs number of distinct keywords in a dataset. 4. Number of docs in a dataset vs time of building index. 5. Number of documents vs time of generating index. 6. Number of keywords in query vs time of generating trapdoors. |
| Keyword matching | Coordinate matching: as many matches as possible. |
| Evaluation parameters | 1. Access pattern. 2. Search Pattern. 3. Server search time. 4. Trapdoor used and its size. 5. Index Size. 6. Ranking Formula. 7. Document Length. 8. No of index words. |
| Security and Consistency | Based on hardness of breaking encryption methods. |

Conjunctive keyword search combines one or more keywords using AND conjunctive. For example, Bangalore AND Mangalore. Fuzzy keyword search matches exact keyword or closely matched keyword in database based on their semantics. Tables 5, 6, 7, 8 and 9 lists some of the known SE schemes, their type i.e. public key or private key, corresponding keyword search types, techniques used, search method, index organization etc.

Table 5: Public Key Searchable Encryption Schemes

| Schemes (Year) | Keyword Search Type | Based on | Performance Measured through |
|---|---|---|---|
| Boneh (2004) | Conjunctive keyword | Bilinear map, Identity Based Encryption (IBE), Decision Diffie-Hellman assumption, General trapdoor | Public key size |
| | | permutation | |
| Golle (2004) | Conjunctive keyword | Decision Diffie-Hellman assumption | Encrypted key hardness |
| Hwang (2007) | Conjunctive keyword | Decision Diffie-Hellman assumption | Shortest private key and cipher text size |
| Boneh (2007) | Conjunctive keyword | Hidden Vector Encryption | Ciphertext Size, Token Size |
| Katz (2008) | Conjunctive keyword | Identity based encryption, Predicate encryption | Secret key and ciphertext security |
| Shi (2007) | Conjunctive keyword | Predicate encryption | Secret key security |
| Chuah (2011) | Fuzzy Keyword Search | Bedtree approach | Storage time |
| Keita (2013) | Oblivious keyword search | Anonymous IBE | Cost of encryption |
| Li (2010) | Fuzzy Keyword Search | Edit distance | Storage cost and searching cost |
| Liu (2011) | Fuzzy Keyword Search | Edit distance | Storage cost, index size, communication cost |

Table 6: Secret Key Searchable Encryption Schemes

| **Schemes** | **Year** | **Keyword Search Type** |
|---|---|---|
| Curtmola | 2006 | Keyword |
| Swaminathan | 2007 | Ranked Keyword |
| Wang | 2010 | Ranked Keyword |
| Zerr | 2008 | Ranked Keyword |
| Yang | 2011 | Keyword |
| Sun | 2013 | Ranked Keyword |
| Chang | 2005 | Keyword |
| Cao | 2010 | Ranked Keyword |
| Bao | 2008 | Keyword |

Table 7: SE Schemes categorized based on search method and type of cryptosystem

| **Search method** | **Secret key Schemes/year** | **Public key Schemes/year** |
|---|---|---|
| **Linear** : Search for each keyword linearly in database | Agrawal/2004, A.Boldyreva/2009, O.Goldreich/1996, D.X.Song/2000. | M.Bellare/2007, T.Fuhr/2007, D.Hofheinz/2008, L. Ibraimi/2011, Q.Tang/2012, |
| **Index based** : Search for each keyword based on index table | F.Bao/2008, C.Bosch/2012, Y.Chang/2005, M.Chase/2010, R.Curtmola/2006, E.J.Goh/2003, P.Golle/2004, B.Hore/2012, M.Islam/2012, F.Kerschbaum/2011, M.Raykova/2011, E.Shen/2009 | J.Baek/2006, J.Baek/2008, D.Boneh/2004, D.Boneh/2007, Y.H.Hwang/2007, J.Katz/2008, H.S.Rhee/2010, Q.Tang/2009, R.Zhang/2007 |

Table 8: Classification of SE methods based on query types

| **Query type** | **Schemes/year** |
|---|---|
| **Equality Test:** A query which searches for equality of keywords. For example rigid is equivalent to hard | R.Agrawal/2004, J.Baek/2006, J.Baek/2008, F.Bao/2008, M.Bellere/2007, A.Boldyreva/2009, D.Boneh/2004, D.Boneh/2007,Y.Chang/2005, M.Chase/2010, R.Curtmola/2006, T.Fuhr/2007, E.J.Goh/2003, O.Goldreich/1996, P.Golle/2004, D.Hofheinz/2008, B.Hore/2012, Y.H.Hwang/2007, L.Ibraimi/2011, V.Iovino/2008, M.Islam/2012, J.Katz/2008,F.Kerschbaum/2011, M.Kuzu/2012, M.Raykova/2012, H.S.Rhee/2010, E.Shen/2009, R.Zhang/2007 |
| **Conjunctive and Disjunctive:** A query with conjunctive or disjunctive operators. For example, search for keywords Bangalore AND Mangalore | D.Boneh/2007, P.Golle/2004, Y.H. Hwang/2007, V.Iovino/2008, J.Katz/2008 |
| **Range:** A range query is a common database operation that retrieves all records where some value is between an upper and lower boundary. For example, list all employees with 3 to 5 years' experience. | R.Agrawal/2004, M.Bellere/2007, A.Boldyreva/2009, D.Boneh/2007, B. Hore/2012, M.Kuzu/2012 |

Table 9: SE schemes based on indexing

| **Schemes based on Index organized by keywords/year** | **Schemes based on Index constructed per document/year** | **Schemes based on Tree based index structure /year** |
|---|---|---|
| Curtmola/2006 | Cao/2010 | Comer/1979 |
| Kamara/2012 | Chang/2005 | Scheuermann/1982 |
| Swaminathan/2007 | Goh/2003 | Lu,y/2012 |
| Wang/2010 | Hwang/2007 | Sun /2013 |

Table 10: Comparison of different SE schemes based on categories

| **Category** | **Schemes Keyword related /Year** | **Conjunction of keywords related schemes/Year** | **Complex queries related schemes/Year** | |
|---|---|---|---|---|
| **Single user:** Cryptographic components are divided between user and the server. Only one user has the key for encrypting, querying and decrypting data. Key is shared between user and server. | Song et al. /2000 , Goh /2003 , Chang & Mitzenmacher/2005, Hacigumus et al. /2005 | Golle et al. /2004, Bosch et al./ 2011 | Hore et al. /2004, Wang & Lakshmanan /2006, Popa et al./2011 | |
| **Semi fledged multi user:** Writing operation on data can be done only by single user, other authorized users can perform read operation on data. Key is shared among users. | Boneh et al./2004, Curtmola et al./2006 , Zhu et al. /2011 | Baek et al. /2008 ,Rhee et al. /2010 ,Cao et al. /2011 | Boneh & Waters/2007, Katz et al. /2008, Yang et al. /2011, Li et al. /2011 ,Lu /2012 | |
| **Full-fledged multi user:** Writing and Reading | Bao et al. /2008 , Dong et al. /2008 , Shao et al. /2010 | Hwang et al./ 2007 | Muhammad Rizwan Asghar et al./2013 | |
| operation on data can be done by all the users without sharing keys. | | | | |

Graph 1: Searchable scheme comparison

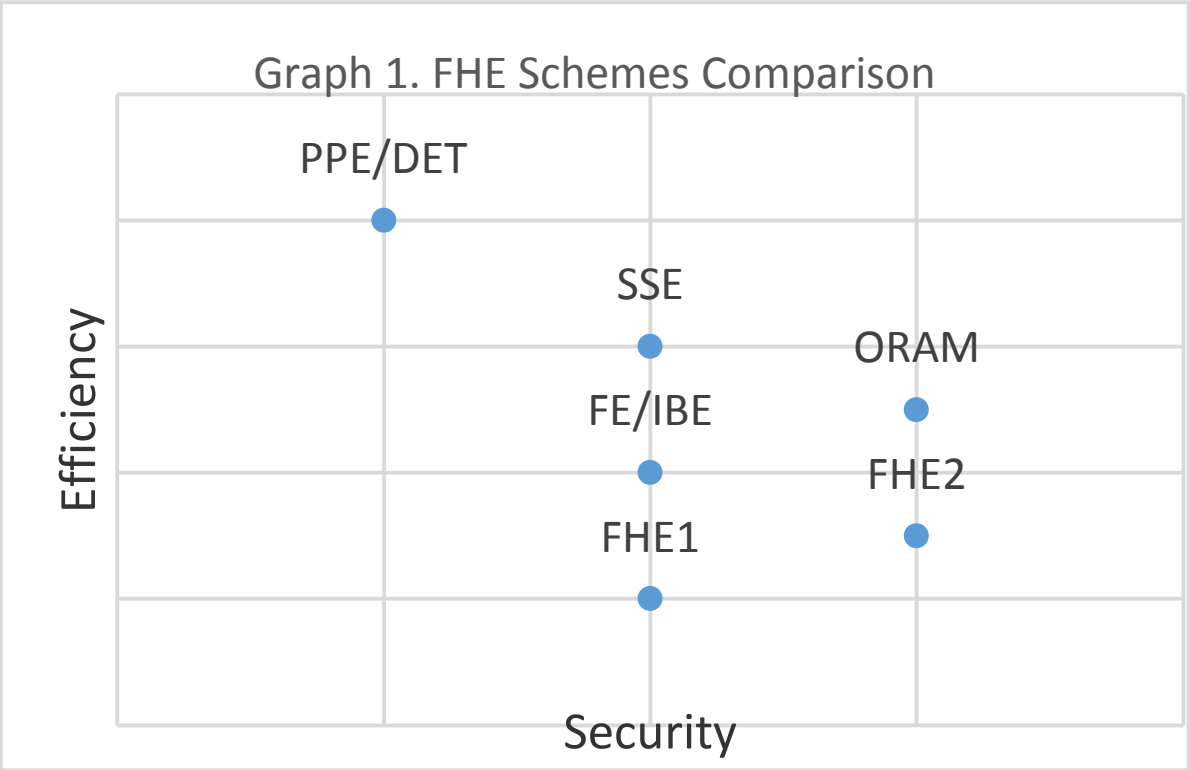

Table 10 compares different schemes based on expressiveness of queries and keywords involving Single user, Semi fledged multi users or Full-fledged multi users [24].

SE schemes can be classified, designed and developed by using many of the above mentioned criteria's or methods. In recent years more research is happening towards developing dynamic searchable encryption schemes with high efficiency.

## IV. Performance

Many techniques like Multi Party Computation, Oblivious RAM (ORAM), Searchable Symmetric encryption (SSE), Searchable Asymmetric Encryption (SAE), Functional Encryption (FE), Property Preserving Encryption (PPE), Homomorphic Encryption (HE), Fully Homomorphic Encryption (FHE), ID based Encryption (IDE), Deterministic Encryption (DET) etc. have been used to perform computation and search on encrypted data.

The Graph 1 compares different algorithms with their efficiency and security. Here the efficiency is taken in terms of performance and security in terms of hardness of breaking encryption methods. PPE is more efficient but less secured. While FHE is less efficient but more secured. SSE falls in between PPE and FHE i.e. SSE efficiency is more compared to FHE but less secured than FHE.

It looks SE is a better tradeoff between security and efficiency. SE does a very fast parallel and I/O efficient keyword search compared to HE, FE, FHE and ORAM. The leaking of certain information by adversaries is the main drawback of SE schemes. Adversaries exploit access patterns and search patterns to leak the information. Hence more research needs to be focused on securing access and search patterns.

## V. Conclusion

Cloud computing is an approach to get hardware or software in a cost effective way. Cloud computing is used as a utility model like electricity, where we pay for what we use. Since Cloud is maintained by third parties, security and privacy plays a vital role in making cloud a popular and successful technology. To provide confidentiality and privacy for our data in cloud we need to concentrate and study different cryptosystems available for use in literature. SE and HE algorithms are most popular cryptosystems used in cloud. HE methods are highly secured but less efficient compared to SE methods, whereas SE methods are more efficient and less secured compared to HE methods. SE is a best tradeoff between efficiency and security. By considering different criteria's of SE methods one can devise a new efficient hybrid algorithm to improve the hardness of security as close as possible to FHE methods.

In future, one can explore efficient SE techniques that work on encrypted data in cloud, for both single user and multi user environment.